\documentclass[twocolumn,prx,aps,amsmath,amssymb,longbibliography]{revtex4-1}

\usepackage{graphicx}
\usepackage{dcolumn}
\usepackage{bm}

\begin{document}

\title{Majorana differential shot noise and its universal thermoelectric
  crossover}

\author{Sergey Smirnov}
\affiliation{P. N. Lebedev Physical Institute of the Russian Academy of
  Sciences, 119991 Moscow, Russia}
\email{1) sergej.physik@gmail.com\\2)
  sergey.smirnov@physik.uni-regensburg.de\\3) ssmirnov@sci.lebedev.ru}

\date{\today}

\begin{abstract}
Nonequilibrium states driven by both electric bias voltages $V$ and
temperature differences $\Delta T$ (or thermal voltages
$eV_T\equiv k_B\Delta T$) are unique probes of various systems. Whereas
average currents $I(V,V_T)$ are traditionally measured in majority of
experiments, an essential part of nonequilibrium dynamics, stored particularly
in fluctuations, remains largely unexplored. Here we focus on Majorana quantum
dot devices, specifically on their differential shot noise
$\partial S^>(V,V_T)/\partial V$, and demonstrate that in contrast to the
differential electric or thermoelectric conductance,
$\partial I(V,V_T)/\partial V$ or $\partial I(V,V_T)/\partial V_T$, it reveals
a crossover from thermoelectric to pure thermal nonequilibrium behavior. It is
shown that this Majorana crossover in $\partial S^>(V,V_T)/\partial V$ is
induced by an interplay of the electric and thermal driving, occurs at an
energy scale determined by the Majorana tunneling amplitude, and exhibits a
number of universal characteristics which may be accessed in solely noise
experiments or in combination with measurements of average currents.
\end{abstract}

\maketitle

\section{Introduction}\label{intro}
Topological superconductors interacting with nanoscopic setups provide a
feasible technological platform to entangle a specific quantum core of these
setups with non-Abelian Majorana bound states (MBSs)
\cite{Kitaev_2001,Alicea_2012,Flensberg_2012,Sato_2016,Aguado_2017,Lutchyn_2018,Valkov_2022}
imitating particle-antiparticle paradigm of the Abelian Majorana fermions
\cite{Majorana_1937} known in the particle physics. Such Majorana entangled
setups are understood (without a strict relation to the specific meaning of
"entanglement" in quantum information) as those where Majorana and
non-Majorana degrees of freedom are coupled via a certain quantum mechanical
mechanism. They represent a special class of condensed matter systems which
are very attractive both from theoretical and experimental perspectives. On
one side, they admit an observation of diverse physical phenomena governed
essentially by Majorana entangled states and, on the other side, they may
function as elementary blocks integrated in various fault tolerant topological
quantum computing \cite{Kitaev_2003} schemes designed to process the quantum
nonlocality supported by MBSs.

Nanoscopic setups where MBSs are entangled with the quantum degrees of freedom
involved in experimental measurements reveal various remarkable
characteristics many of which may be accessed in quantum transport
experiments. Such experiments deal with nonequilibrium states which may be
generated by bias voltages $V$ or temperature differences $\Delta T$,
expressed equivalently through the corresponding thermal voltages $V_T$,
defined as $eV_T\equiv k_B\Delta T$. Here Majorana features are predicted
within the framework of mean currents $I(V)$ induced by mostly bias voltages
\cite{Liu_2011,Fidkowski_2012,Prada_2012,Pientka_2012,Lin_2012,Lee_2013,Kundu_2013,
  Vernek_2014,Ilan_2014,Cheng_2014,Lobos_2015,Peng_2015,Sharma_2016,Heck_2016,
  Das_Sarma_2016,Lutchyn_2017,Liu_2017,Huang_2017,Liu_2018,Lai_2019,Tang_2020,
  Zhang_2020,Chi_2021,Wang_2021,Galambos_2022,Jin_2022,Zou_2023}
or, to a lesser extent, mean currents $I(V,V_T)$ induced by also temperature
differences
\cite{Leijnse_2014,Lopez_2014,Khim_2015,Ramos-Andrade_2016,Smirnov_2020a,He_2021,
  Giuliano_2022,Buccheri_2022,Majek_2022,Bondyopadhaya_2022,Zou_2022}.
Experimental efforts on mean currents $I(V)$ induced by bias voltages
\cite{Mourik_2012,Nadj-Perge_2014,Wang_2022} are aimed to measure the
differential conductance $\partial I(V)/\partial V$. Of particular interest
here is the zero bias limit of the differential conductance (linear
conductance) which should attain a certain universal value predicted
theoretically for a specific Majorana entangled setup. Although such mean
current experiments are well developed and should be performed in the first
place, unfortunately, they may be controversial \cite{Yu_2021,Frolov_2021} in
detecting MBSs and, as a consequence, other types of quantum transport
measurements, or perhaps sequences of measurements \cite{Ziesen_2023}, are
currently in demand. Particularly, one is interested in those physical
observables which demonstrate in a given setup a Majorana driven behavior
which is qualitatively different from the behavior of the mean currents
measured in the same Majorana entangled setup.

An attractive quantum transport alternative to the mean value of a current
flowing through a nanoscopic setup is to study the random deviations of this
current from its mean value, that is the current fluctuations, characterized,
for example, by the shot noise $S^>$. Here majority of Majorana shot noise
proposals assume nonequilibrium states originating from bias voltages
\cite{Liu_2015,Liu_2015a,Haim_2015,Valentini_2016,Zazunov_2016,Smirnov_2017,
  Jonckheere_2019,Smirnov_2019,Manousakis_2020,Feng_2022,Smirnov_2022} and
explore the behavior of $S^>(V)$ at small and large $V$. As in mean current
experiments, where the differential conductance $\partial I(V)/\partial V$
provides an access to an averaged Majorana universality, the differential shot
noise $\partial S^>(V)/\partial V$ allows one to reveal a universal
fluctuation behavior governed by Majorana entangled states.

One may also avoid resorting to nonequilibrium behavior and address Majorana
entangled states in corresponding equilibrium nanoscopic setups by means of
quantum thermodynamic tools such as the entropy of these setups
\cite{Smirnov_2015,Sela_2019,Smirnov_2021,Smirnov_2021a,Ahari_2021}. Recent
experimental and theoretical activities
\cite{Hartman_2018,Kleeorin_2019,Pyurbeeva_2021,Child_2022,Han_2022,Pyurbeeva_2022,Child_2022a}
on the entropy of nanoscale and mesoscale systems demonstrate that this
engrossing approach may become a powerful and univocal technique which will
not be subject to further controversy similar to the one about the Majorana
differential conductance.

Nevertheless, presently quantum transport is a more appealing framework within
which experimentalists have at their disposal well established technologies
verified in diverse nanoscopic setups for a long period of time. Moreover,
quantum transport techniques have a wider space of control due to numerous
additional parameters utilized to maintain various kinds of nonequilibrium
states in which a broad spectrum of physical observables is available for
performing experiments. Thus applying the quantum transport framework to
Majorana entangled nanoscopic setups provides vast freedom in exploring
Majorana phenomena in nonequilibrium. In particular, among measurements of
other physical observables, shot noise experiments in various nonequilibrium
states are expected to qualitatively enrich the existing results on mean
currents in Majorana entangled nanoscopic setups.

Here we focus on the shot noise in nonequilibrium states produced by bias
voltages $V$ and thermal voltages $V_T$ in a quantum dot (QD) whose degrees of
freedom are entangled with MBSs of a topological superconductor. Specifically,
we explore the differential shot noise $\partial S^>(V,V_T)/\partial V$ which,
as has been discussed above, inspects universal Majorana fluctuation
behavior. So far it is not much known about this physical observable when both
$V$ and $V_T$ excite competing current flows in a Majorana setup. Indeed,
whereas the differential thermoelectric shot and quantum noise,
$\partial S^>(V,V_T)/\partial V_T$, have been addressed
\cite{Smirnov_2018,Smirnov_2019a} in presence of both bias voltages and
temperature differences, the differential shot noise
$\partial S^>(V,V_T)/\partial V$ remains to a large extent unexplored for
Majorana entangled setups in nonequilibrium states driven by both $V$ and
$V_T$. It should be noted that in nonequilibrium states induced only by bias
voltages $V$ the differential shot noise has been studied in combination with
the differential conductance. In particular, in Ref. \cite{Cao_2023} it is
demonstrated that in presence of MBSs a dip of the differential shot noise is
always accompanied by a peak of the differential conductance. This behavior
has also been observed earlier in Ref. \cite{Smirnov_2022} (see its Fig. 4,
namely, the insets of the upper panel). As mentioned above, it is important to
find for a given Majorana entangled setup physical observables whose behavior
has a character qualitatively different from the one of the mean current or
its derivative physical quantities such as the differential conductance
$\partial I(V,V_T)/\partial V$ or differential thermoelectric conductance
$\partial I(V,V_T)/\partial V_T$ whose behavior may be obtained in the same
setup. We demonstrate that the differential shot noise is one of such physical
observables which is distinguished by the presence of a crossover from a
thermoelectric to pure thermal nonequilibrium behavior. It is shown that
whereas the differential shot noise passes through its crossover, the
differential conductance and differential thermoelectric conductance do not
exhibit any crossover or any other peculiarity. Thus, in contrast to
$\partial I(V,V_T)/\partial V$ and $\partial I(V,V_T)/\partial V_T$, the
differential shot noise $\partial S^>(V,V_T)/\partial V$ brings a
nonequilibrium energy scale having a pure fluctuation nature meaning that it
cannot be revealed within measurements limited only by the mean
current. Besides being of fundamental interest, the energy scale associated
with the nonequilibrium crossover in $\partial S^>(V,V_T)/\partial V$ is shown
to be of practical importance in expressing quantitatively the fluctuation
universality of Majorana entangled states via a number of measurable universal
ratios which would be of interest for future experiments.

The paper is organized as follows. In Section \ref{meqd} we discuss a
theoretical model of an experimentally feasible nanoscopic setup where MBSs
are entangled with a QD whose nonequilibrium states are generated by both a
bias voltage and thermal voltage. Results of numerical analysis performed with
high accuracy for the differential shot noise
$\partial S^>(V,V_T)/\partial V$, differential conductance
$\partial I(V,V_T)/\partial V$ and differential thermoelectric conductance
$\partial I(V,V_T)/\partial V_T$ are presented in Section \ref{nadsn} where it
is demonstrated that, in contrast to $\partial I(V,V_T)/\partial V$ and
$\partial I(V,V_T)/\partial V_T$, one observes in $\partial
S^>(V,V_T)/\partial V$ a crossover from a thermoelectric to pure thermal
nonequilibrium behavior. The energy scale where this crossover takes place and
a number of universal Majorana ratios involving this energy scale are also
shown in this section. Finally, with Section \ref{concl} we make conclusions
and discuss possible outlooks.
\section{Theoretical model of a Majorana entangled quantum dot and the
  differential shot noise in thermoelectric nonequilibrium}\label{meqd}
We start with a description of a setup which, on one side, is technologically
feasible \cite{Deng_2016,Deng_2018} and, on the other side, involves a basic
mechanism of a Majorana entanglement which is sufficient to demonstrate a
remarkable nonequilibrium behavior of the differential shot noise in presence
of both bias voltages and temperature differences. To this end, let us
consider a noninteracting QD,
\begin{equation}
  \hat{H}_{QD}=\epsilon_d d^\dagger d.
  \label{Ham_QD}
\end{equation}
The QD is nondegenerate and its energy level $\epsilon_d$ is tunable by a
proper gate voltage. The choice of a setup with a noninteracting QD is quite a
realistic assumption to explore universal Majorana phenomena at low
energies. Indeed, the spin degeneracy is assumed to be removed by an external
magnetic field which excludes a possible interfering of an interaction induced
Kondo universal behavior, well-known in experiment and theory of
spin-degenerate QDs
\cite{Ralph_1994,Goldhaber-Gordon_1998,Meir_1993,Wingreen_1994,Smirnov_2011a,Smirnov_2011b,Niklas_2016},
with the low-energy Majorana universal effects, which are of interest in this
work. Numerical renormalization group calculations \cite{Tijerina_2015} have
demonstrated that interacting spin-degenerate QDs in external magnetic fields
behave similar to noninteracting nondegenerate QDs exhibiting, for example,
the linear conductance $e^2/2h$ which results entirely from the Majorana
entangled states. Thus Eq. (\ref{Ham_QD}) is a proper model to explore
low-energy Majorana quantum transport, in particular, in nonequilibrium states
resulting from bias voltages and temperature differences \cite{Lopez_2014}.

Two normal noninteracting metallic contacts, denoted below as left ($L$) and
right ($R$),
\begin{equation}
  \hat{H}_C=\sum_{l=\{L,R\}}\sum_k\epsilon_kc_{lk}^\dagger c_{lk},
  \label{Ham_C}
\end{equation}
are connected to the QD via tunneling processes,
\begin{equation}
  \hat{H}_{QD-C}=\sum_{l=\{L,R\}}\mathcal{T}_l\sum_kc_{lk}^\dagger d+\text{H.c.}
  \label{Ham_C-QD}
\end{equation}
In Eq. (\ref{Ham_C}) the continuous energy spectrum $\epsilon_k$ gives rise to
a density of states of the contacts $\nu_C(\epsilon)$ which is in general
energy dependent. However, around the Fermi energy one usually with a good
accuracy assumes that its energy dependence plays no essential role for
quantum transport and thus $\nu_C(\epsilon)\approx\nu_0/2$. In
Eq. (\ref{Ham_C-QD}) one additionally assumes that the tunneling matrix
elements do not depend on the set $k$ of the quantum numbers used to describe
the states in the contacts, $\mathcal{T}_{kl}\approx\mathcal{T}_l$. The
relevant energy scales brought about by the tunneling between the QD and
contacts are $\Gamma_l=\pi\nu_0|\mathcal{T}_l|^2$. By proper gate voltages one
may achieve the symmetric coupling $\Gamma_L=\Gamma_R=\Gamma/2$ which will be
assumed below.

Each contact is assumed to be in its own equilibrium state with the
corresponding Fermi-Dirac distribution,
\begin{equation}
  f_l(\epsilon)=\frac{1}{\exp\bigl(\frac{\epsilon-\mu_l}{k_BT_l}\bigr)+1}.
  \label{FD}
\end{equation}
Here the chemical potentials,
\begin{equation}
  \mu_{L,R}=\mu_0\pm eV/2,
  \label{Chem_pot}
\end{equation}
are specified by the bias voltage $V$ such that $eV<0$ and the temperature of
the left contact is higher than the temperature of the right contact, that is
\begin{equation}
  \begin{split}
    &T_L=T+\Delta T\quad\text{(hot contact)},\\
    &T_R=T\quad\text{(cold contact)},
  \end{split}
  \label{Temp}
\end{equation}
assuming $T,\Delta T\geqslant 0$. The QD is out of equilibrium when either
$V\neq 0$ or $\Delta T\neq 0$.

A topological superconductor hosting two MBSs $\gamma_{1,2}$ at its ends,
\begin{equation}
  \hat{H}_{TS}=\frac{1}{2}i\xi\gamma_2\gamma_1,\quad\gamma_{1,2}^\dagger=\gamma_{1,2},\quad\{\gamma_i,\gamma_j\}=2\delta_{ij},
  \label{Ham_TS}
\end{equation}
interacts with the QD,
\begin{equation}
  \hat{H}_{QD-TS}=\eta^*d^\dagger\gamma_1+\text{H.c.},
  \label{Ham_TS-QD}
\end{equation}
implementing a direct entanglement of the QD's degrees of freedom with the
Majorana mode $\gamma_1$ of the topological superconductor. In
Eq. (\ref{Ham_TS}) the parameter $\xi$ is an energetic measure of how strong
the two Majorana modes overlap with each other. When $\xi$ is small the MBSs
are well separated whereas large values of $\xi$ model a situation where the
two MBSs merge into a single Dirac fermion. In Eq. (\ref{Ham_TS-QD}) the
Majorana tunneling amplitude $|\eta|$ specifies the strength of the Majorana
entanglement.

A schematic summary of the above theoretical formulation of the setup, based
on Eqs. (\ref{Ham_QD})-(\ref{Ham_TS-QD}), is illustrated in the inset of
Fig. \ref{figure_1}.

The Hamiltonian of the setup,
$\hat{H}=\hat{H}_{QD}+\hat{H}_C+\hat{H}_{QD-C}+\hat{H}_{TS}+\hat{H}_{QD-TS}$,
allows us to formulate the problem in terms of the Keldysh field integral
\cite{Altland_2010}, a convenient tool to calculate various correlation
functions. Other technical tools based, {\it e.g.}, on quantum master
equations \cite{Xu_2022} may also be considered as alternative approaches to
the problem. Within the Keldysh field integral formalism one may
straightforwardly derive the shot noise from the Keldysh generating
functional,
\begin{equation}
  Z[J_{lq}(t)]=\int\mathcal{D}[\bar{\psi},\psi;\bar{\phi},\phi;\bar{\zeta},\zeta]e^{\frac{i}{\hbar}S_K[J_{lq}(t)]},
  \label{Z_K}
\end{equation}
which is a field integral over the Grassmann fields of the QD
($\bar{\psi}_q(t),\psi_q(t)$), contacts ($\bar{\phi}_{lkq}(t),\phi_{lkq}(t)$)
and topological superconductor ($\bar{\zeta}_q(t),\zeta_q(t)$) whose temporal
arguments are on the real axis and $q=\pm$ specifies, respectively, the
forward or backward branch of the Keldysh contour. At zero source fields the
Keldysh generating functional is determined by the Keldysh action
$S_K^{(0)}\equiv S_K[J_{lq}(t)=0]$ and is equal to unity,
$Z[J_{lq}(t)=0]=1$. The Keldysh action $S_K[J_{lq}(t)]$,
\begin{equation}
  \begin{split}
    &S_K[J_{lq}(t)]=S_{QD}[\bar{\psi},\psi]+S_C[\bar{\phi},\phi]+S_{TS}[\bar{\zeta},\zeta]\\
    &+S_{QD-C}[\bar{\psi},\psi;\bar{\phi},\phi]+S_{QD-TS}[\bar{\psi},\psi;\bar{\zeta},\zeta]\\
    &+S_O[\bar{\psi},\psi;\bar{\phi},\phi;J_{lq}(t)],
  \end{split}
  \label{S_K}
\end{equation}
is the sum of, respectively, the actions describing the QD, contacts,
topological superconductor, tunneling between the QD and contacts, tunneling
between the QD and topological superconductor and the source action added
to generate an observable of interest, in particular, the mean current and
shot noise. The actions $S_{QD}$, $S_C$ and $S_{TS}$ are of the standard
matrix form \cite{Altland_2010} in the retarded-advanced space. The actions
$S_{QD-C}$, $S_{QD-TS}$ and $S_O$ have the following form:
\begin{equation}
  S_{QD-C}=-\!\int_{-\infty}^\infty \!\!\!\!\!dt\sum_{l=\{L,R\}}\sum_{k,q}[\mathcal{T}_lq\bar{\phi}_{lkq}(t)\psi_q(t)+\text{G.c.}],
  \label{S_QD-C}
\end{equation}
\begin{equation}
  \begin{split}
    S_{QD-TS}&=-\int_{-\infty}^\infty dt\{\eta^*\sum_qq[\bar{\psi}_q(t)\zeta_q(t)\\
    &+\bar{\psi}_q(t)\bar{\zeta}_q(t)]+\text{G.c.}\},
  \end{split}
  \label{S_QD-TS}
\end{equation}
\begin{equation}
  S_O=-\int_{-\infty}^\infty dt\sum_{l=\{L,R\}}\sum_qJ_{lq}(t)I_{lq}(t),
  \label{S_O}
\end{equation}
where $\text{G.c.}$ denotes the Grassmann conjugated terms and $I_{lq}(t)$ is
the current operator in the Grassmann representation,
\begin{equation}
  I_{lq}(t)=\frac{ie}{\hbar}\sum_k\bigl(\mathcal{T}_l\bar{\phi}_{lkq}(t)\psi_q(t)-\text{G.c.}\bigr).
  \label{Current}
\end{equation}
The form of the source action in Eq. (\ref{S_O}) implies that one derives the
mean current and current-current correlations via proper differentiations,
\begin{equation}
  \langle I_{lq}(t)\rangle_0=i\hbar\frac{\delta Z[J_{lq}(t)]}{\delta J_{lq}(t)}\biggl|_{J_{lq}(t)=0},
  \label{Diff_mean_current}
\end{equation}
\begin{equation}
  \langle I_{lq}(t)I_{l'q'}(t')\rangle_0=(i\hbar)^2\frac{\delta^2 Z[J_{lq}(t)]}{\delta J_{lq}(t)\delta J_{l'q'}(t')}\biggl|_{J_{lq}(t)=0},
  \label{Diff_current-current}
\end{equation}
where
\begin{equation}
  \begin{split}
    &\langle\prod_iI_{l_iq_i}(t_i)\rangle_0\\
    &\equiv\int\mathcal{D}[\bar{\psi},\psi;\bar{\phi},\phi;\bar{\zeta},\zeta]e^{\frac{i}{\hbar}S_K^{(0)}}\prod_iI_{l_iq_i}(t_i).
  \end{split}
  \label{Average_0}
\end{equation}

Choosing the left contact as the one where measurements of the mean current
\begin{equation}
  I(V,V_T)=\langle I_{Lq}(t)\rangle_0,
  \label{Mean_current_left}
\end{equation}
and correlations
\begin{equation}
  S^>(t,t';V,V_T)=\langle\delta I_{L-}(t)\delta I_{L+}(t')\rangle_0
  \label{Current_fluct_corr_left}
\end{equation}
of the current fluctuations
\begin{equation}
  \delta I_{Lq}(t)=I_{Lq}(t)-I(V,V_T)
  \label{Current_fluct_left}
\end{equation}
are performed, one obtains the shot noise $S^>(V,V_T)$ in the left contact as
the zero frequency Fourier transform of $S^>(t,t';V,V_T)=S^>(t-t';V,V_T)$,
\begin{equation}
  \begin{split}
    &S^>(\omega;V,V_T)=\int_{-\infty}^\infty dt\,e^{i\omega t}S^>(t;V,V_T),\\
    &S^>(V,V_T)=S^>(\omega=0;V,V_T).
  \end{split}
  \label{Shot_noise}
\end{equation}

As it has already been mentioned in Section \ref{intro}, majority of quantum
transport experiments deal with mean currents, Eq. (\ref{Mean_current_left}),
specifically, with their differential characteristics such as the differential
conductance or, less often, differential thermoelectric conductance,
$\partial I(V,V_T)/\partial V$ or $\partial I(V,V_T)/\partial V_T$,
respectively. These quantities have universal units of $e^2/h$ and thus
provide direct access to universal properties of MBSs. Likewise, experiments
dealing with shot noises, Eq. (\ref{Shot_noise}), and their derivatives, may
access universal fluctuation behavior of Majorana entangled states via, for
example, the differential shot noise, $\partial S^>(V,V_T)/\partial V$, having
universal units of $e^3/h$. Whereas the Majorana universality of
$\partial S^>/\partial V$ is still an experimental challenge for Majorana
entangled setups, the differential shot noise has already been successfully
measured to probe other types of fluctuation universality, for example the
universality of the Kondo noise in quantum dots \cite{Basset_2012}. Although
experiments on current fluctuations are more complicated than those measuring
mean currents, results of such noise measurements provide a much more detailed
microscopic structure of various nanoscopic setups.

Below we obtain the differential conductance $\partial I(V,V_T)/\partial V$,
differential thermoelectric conductance $\partial I(V,V_T)/\partial V_T$ and
differential shot noise $\partial S^>(V,V_T)/\partial V$ by means of numerical
calculations based on finite differences used to approximate the corresponding
derivatives. Here we would like to emphasize that although the above
theoretical model is noninteracting, numerical calculations of $S^>(V,V_T)$
and $I(V,V_T)$ are nevertheless necessary. The point is that after obtaining
closed analytic expressions for $S^>(V,V_T)$ and $I(V,V_T)$, which is possible
because the Keldysh field integral is quadratic in the fermionic fields, it
still remains to perform integrals in the energy domain (see the Appendix in
Ref. \cite{Smirnov_2018}) in these analytic expressions. These integrals are
hard to calculate analytically, especially, at finite temperature differences
(characterized by finite thermal voltages $V_T$), that is when the Fermi-Dirac
distributions in Eq. (\ref{FD}) cannot be approximated by step-like
functions. In general, calculations of the differential shot noise are more
time consuming than those which would be necessary to get just the shot
noise. Whereas a certain numerical accuracy may be sufficient to get curves
looking smooth enough for $S^>(V,V_T)$, using the same numerical data to
calculate $\partial S^>(V,V_T)/\partial V$ may result in numerical errors
leading to a chaotic dataset. Thus to obtain $\partial S^>(V,V_T)/\partial V$
with an accuracy that allows to identify dependence on various parameters as
well as corresponding coefficients, the calculation of $S^>(V,V_T)$ should be
done with a proper precision. Clearly, a higher degree of numerical accuracy
leads to a notable increase of the computational time but still makes it
possible to perform for the setup described in this section a detailed
analysis of $\partial S^>(V,V_T)/\partial V$, in particular, its universal
Majorana thermoelectric crossover discussed thoroughly in the next section.
\section{Numerical analysis of the differential shot noise and its
  thermoelectric crossover}\label{nadsn}
In this section we present numerical results for the differential shot noise
$\partial S^>(V,V_T)/\partial V$ and demonstrate that, in contrast to the
differential conductance $\partial I(V,V_T)/\partial V$ and differential
thermoelectric conductance $\partial I(V,V_T)/\partial V_T$, it exhibits a
crossover from one type of nonequilibrium behavior to a qualitatively
different one. This crossover occurs in the regime
\begin{equation}
  \Gamma\gg eV_T\gg|eV|\gg\xi,
  \label{Quantum_transport_regime}
\end{equation}
and in the most part of this section we focus on quantum transport in this
regime except for the last part where we show that the crossover disappears
for large values of the Majorana overlap energy $\xi$.

\begin{figure}
\includegraphics[width=8.0 cm]{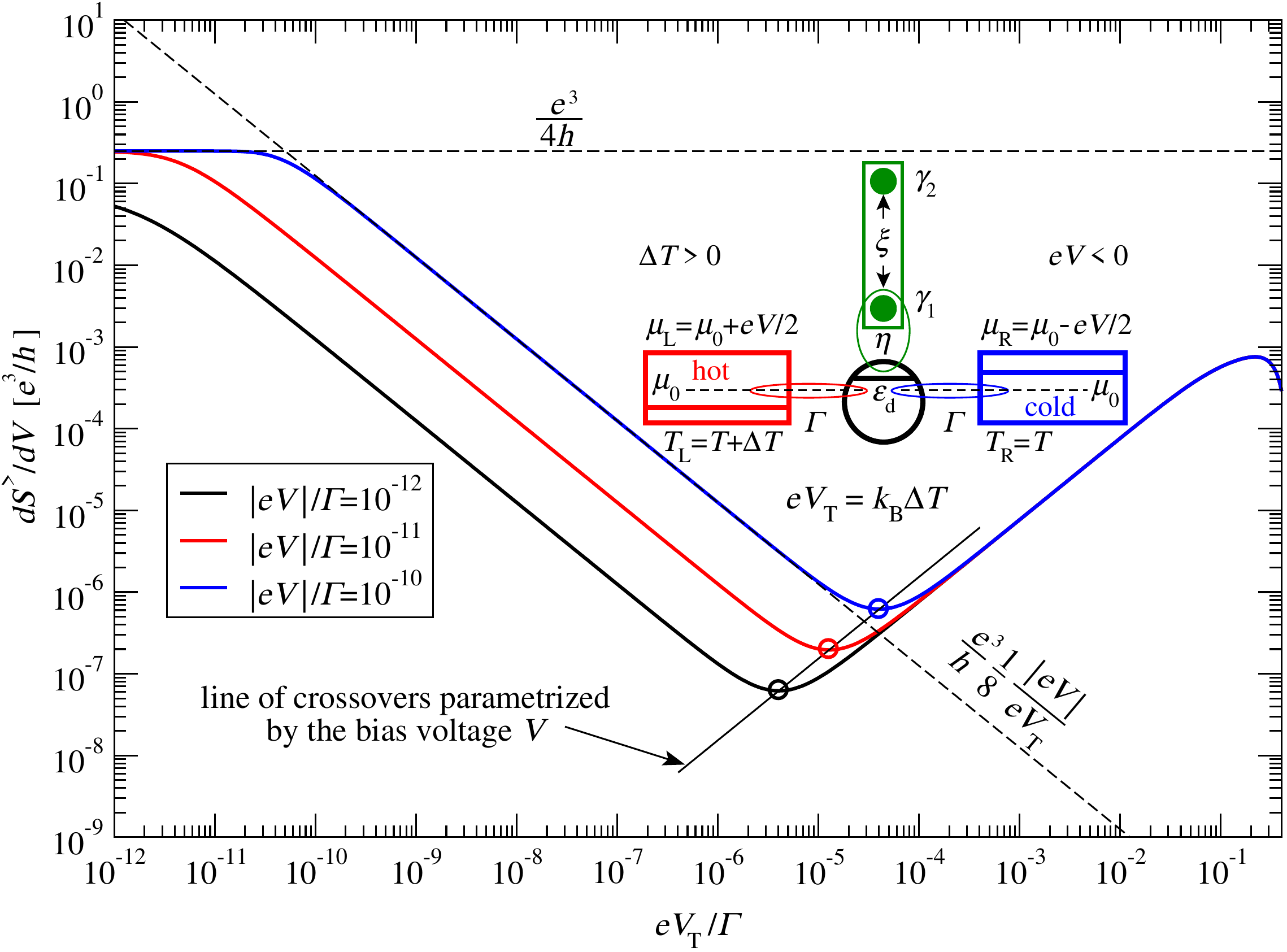}
\caption{\label{figure_1} Differential shot noise $\partial S^>/\partial V$ as
  a function of the thermal voltage $V_T$ for three different values of the
  bias voltage: $|eV|/\Gamma=10^{-12}$ (black), $|eV|/\Gamma=10^{-11}$ (red),
  $|eV|/\Gamma=10^{-10}$ (blue). The other parameters have the following
  values: $\epsilon_d/\Gamma=10^{-1}$, $k_BT/\Gamma=10^{-12}$,
  $|\eta|/\Gamma=1$, $\xi/\Gamma=10^{-14}$. The Majorana device presented above the curves illustrates
  schematically the physical setup, described in the main text,
  Eqs. (\ref{Ham_QD})-(\ref{Ham_TS-QD}), assuming $eV<0$ and $\Delta T>0$.}
\end{figure}

In Fig. \ref{figure_1} we show numerical results obtained for the differential
shot noise $\partial S^>(V,V_T)/\partial V$ as a function of the thermal
voltage $V_T$ for different values of the bias voltage $V$. As can be seen, at
a certain value of the thermal voltage $V_T=V_{T,\text{min}}$ each of the
three curves has a characteristic minimum (shown by the corresponding circle)
which represents a crossover from a thermoelectric to pure thermal
nonequilibrium behavior. Indeed, the decreasing, or thermoelectric, branch
depends on both the electric driving $V$ and thermal driving $V_T$ with the
asymptotic behavior shown by the inclined dashed line whereas the increasing,
or pure thermal, branch depends only on the thermal driving $V_T$ and does not
depend on the electric driving $V$. Our numerical analysis shows that the
asymptotics of the thermoelectric and pure thermal nonequilibrium branches
are, respectively, given by the following analytic expressions:
\begin{equation}
  \begin{split}
    &\frac{\partial S^>(V,V_T)}{\partial V}=\frac{e^3}{h}\frac{1}{8}\frac{|eV|}{eV_T},\\
    &\text{for $V_T$:}\quad|eV|\ll eV_T\ll eV_{T,\text{min}},
  \end{split}
  \label{Thermoelectric_branch}
\end{equation}
and
\begin{equation}
  \begin{split}
    &\frac{\partial S^>(V,V_T)}{\partial V}=\frac{e^3}{h}\frac{1-\ln(2)}{4}\frac{\epsilon_d(eV_T)}{\eta^2},\\
    &\text{for $V_T$:}\quad eV_{T,\text{min}}\ll eV_T\ll\Gamma,
  \end{split}
  \label{Pure_thermal_branch}
\end{equation}
both of which we are able to reproduce with any desired numerical
accuracy. Similarly, in the whole range of the thermal voltage $V_T$,
restricted by the regime specified in Eq. (\ref{Quantum_transport_regime}),
our numerical calculations show that the analytic expression for the
differential shot noise is given by the sum of
Eqs. (\ref{Thermoelectric_branch}) and (\ref{Pure_thermal_branch}),
\begin{equation}
  \begin{split}
    &\frac{\partial S^>(V,V_T)}{\partial V}=\frac{e^3}{h}\biggl[\frac{1}{8}\frac{|eV|}{eV_T}+\frac{1-\ln(2)}{4}\frac{\epsilon_d(eV_T)}{|\eta|^2}\biggr],\\
    &\text{for $V_T$:}\quad|eV|\ll eV_T\ll\Gamma.
  \end{split}
  \label{Diff_SN_whole_range}
\end{equation}
An analytic derivation of Eq. (\ref{Diff_SN_whole_range}) is a complicated
task which we would like to address in a separate paper. We note that a proper
analytic analysis may provide corrections to Eq. (\ref{Diff_SN_whole_range})
and show under which conditions these corrections start to play an essential
role. Also using the Sommerfeld expansion \cite{Ashkroft_1976}, one may
analytically derive the differential shot noise in the complementary regime
where $eV_T\ll|eV|$. However, within the specified regime,
Eq. (\ref{Quantum_transport_regime}), the analytic expression in
Eq. (\ref{Diff_SN_whole_range}) has been confirmed with any desired numerical
precision. This means that the stronger the inequalities in
Eq. (\ref{Quantum_transport_regime}) are fulfilled, the more digits after the
decimal point are reproduced numerically for any given value obtained
analytically from Eq. (\ref{Diff_SN_whole_range}).

From Eq. (\ref{Diff_SN_whole_range}) we find that $eV_{T,\text{min}}$ depends
on the bias voltage $V$, gate voltage $\epsilon_d$ and Majorana tunneling
amplitude $|\eta|$,
\begin{equation}
  eV_{T,\text{min}}=\biggl\{\frac{1}{2[1-\ln(2)]}\frac{|eV||\eta|^2}{\epsilon_d}\biggr\}^\frac{1}{2}.
  \label{V_T_min}
\end{equation}
The differential shot noise at $V_T=V_{T,\text{min}}$ is obtained from
Eqs. (\ref{Diff_SN_whole_range}) and (\ref{V_T_min}) which lead to the
following result:
\begin{equation}
  \frac{\partial S^>(V,V_T)}{\partial V}\bigg|_{V_T=V_{T,\text{min}}}=\frac{e^3}{h}\biggl[\frac{1-\ln(2)}{8}\frac{\epsilon_d|eV|}{|\eta|^2}\biggr]^\frac{1}{2}.
  \label{Diff_SN_V_T_min}
\end{equation}
\begin{figure}
\includegraphics[width=8.0 cm]{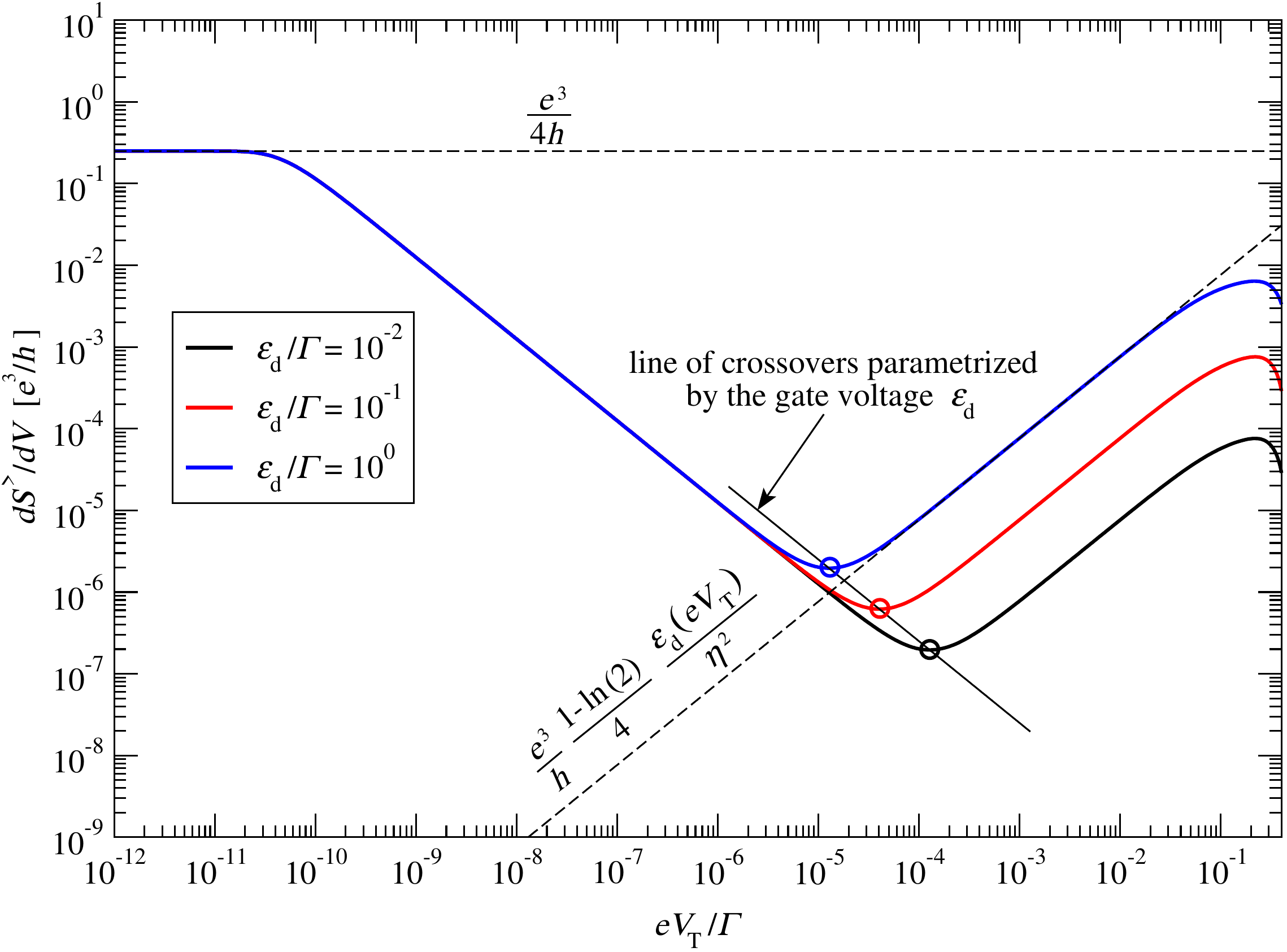}
\caption{\label{figure_2} Differential shot noise $\partial S^>/\partial V$ as
  a function of the thermal voltage $V_T$ for three different values of the
  gate voltage: $\epsilon_d/\Gamma=10^{-2}$ (black),
  $\epsilon_d/\Gamma=10^{-1}$ (red), $\epsilon_d/\Gamma=1$ (blue). The other
  parameters have the following values: $|eV|/\Gamma=10^{-10}$,
  $k_BT/\Gamma=10^{-12}$, $|\eta|/\Gamma=1$, $\xi/\Gamma=10^{-14}$.}
\end{figure}
Note, that the thermoelectric branch, Eq. (\ref{Thermoelectric_branch}), is
universal because it does not depend on the gate voltage $\epsilon_d$ and
depends only on the ratio between the electric and thermal driving, $V$ and
$V_T$, respectively.

According to Eqs. (\ref{V_T_min}) and (\ref{Diff_SN_V_T_min}) both the
location of the crossover, $V_{T,\text{min}}$, and the value of
$\partial S^>(V,V_T)/\partial V$ at $V_T=V_{T,\text{min}}$ depend on $V$. This
suggests that the crossover results from an interplay between the electric and
thermal driving. Moreover, the dependence of the crossover on $\epsilon_d$ in
Eqs. (\ref{V_T_min}) and (\ref{Diff_SN_V_T_min}) (see also
Fig. \ref{figure_2}) is another indication that the crossover emerges from a
competition of the two flows excited, respectively, by the electric and
thermal driving. Obviously, due to the particle-hole symmetry, the current
cannot be excited exclusively by the thermal driving $V_T$ when
$\epsilon_d=0$. The pure thermal driving $V_T$ induces a finite current only
when $\epsilon_d\neq 0$.

The straight solid line in Fig. \ref{figure_1} shows both the locations
$V_{T,\text{min}}$ of the crossovers and the values of
$\partial S^>(V,V_T)/\partial V$ at $V_T=V_{T,\text{min}}$ parameterized by
the bias voltage $V$ according to Eqs. (\ref{V_T_min}) and
(\ref{Diff_SN_V_T_min}). The figure clearly shows that the crossovers
(highlighted by the circles), obtained from the numerical calculations, reside
perfectly on the analytic straight solid line. As expected, at low energies,
$eV_T\ll|eV|$ (that is outside the regime in
Eq. (\ref{Quantum_transport_regime})), the differential shot noise reaches its
universal asymptotic unitary value $e^3/4h$ shown by the horizontal dashed
line (see also Refs. \cite{Liu_2015,Smirnov_2017}).

The numerical results presented in Fig. \ref{figure_2} show the differential
shot noise $\partial S^>(V,V_T)/\partial V$ as a function of the thermal
voltage $V_T$ for different values of the gate voltage $\epsilon_d$. As in
Fig. \ref{figure_1}, each of the three curves possesses a crossover (shown by
the corresponding circle) from a thermoelectric to pure thermal nonequilibrium
behavior. As mentioned above, on the left side of the crossover the
decreasing, or thermoelectric, branch is universal: it depends on both the
electric driving $V$ and thermal driving $V_T$ via their ratio and
Fig. \ref{figure_2} explicitly demonstrates that it does not depend on the
gate voltage $\epsilon_d$. On the right side of the crossover the increasing,
or pure thermal, branch, which is driven only by the thermal voltage $V_T$, is
obviously not universal. Indeed, the figure clearly shows that this branch
depends on the gate voltage $\epsilon_d$. The inclined dashed line shows the
asymptotic behavior of the pure thermal nonequilibrium branch, in particular,
its dependence on the gate voltage $\epsilon_d$. Note, that in fact this pure
thermal nonequilibrium branch is not universal because of two reasons. The
first reason, the dependence on $\epsilon_d$, was already mentioned above. The
second reason is that this branch additionally depends on the Majorana
tunneling amplitude $|\eta|$ as it is also shown in its asymptotic
behavior.

In addition to the dependence on $V$, both the location of the crossover,
$V_{T,\text{min}}$, and the value of $\partial S^>(V,V_T)/\partial V$ at
$V_T=V_{T,\text{min}}$ depend on $\epsilon_d$. The straight solid line shows
both the locations $V_{T,\text{min}}$ of the crossovers and the values of
$\partial S^>(V,V_T)/\partial V$ at $V_T=V_{T,\text{min}}$ parameterized by
the gate voltage $\epsilon_d$ according to Eqs. (\ref{V_T_min}) and
(\ref{Diff_SN_V_T_min}). As in the case of the parametric dependence on $V$,
one clearly sees that the numerically obtained crossovers, marked by the
circles, also reside perfectly on the analytic straight line resulting from
the parametric dependence on $\epsilon_d$. Here the universality of the
differential shot noise at low energies, $eV_T\ll|eV|$, is explicitly visible:
the asymptotic low-energy behavior is obviously independent of $\epsilon_d$
and is characterized by the unitary value $e^3/4h$ shown by the horizontal
dashed line.

The nonequilibrium Majorana crossover in the differential shot noise has a
number of universal properties which may quantitatively be expressed via a
number of ratios taking universal values. For example, according to
Eqs. (\ref{V_T_min}) and (\ref{Diff_SN_V_T_min}) the ratio
\begin{equation}
  R_1\equiv\frac{eV_{T,\text{min}}}{|eV|}\frac{\partial S^>(V,V_T)}{\partial V}\bigg|_{V_T=V_{T,\text{min}}}
  \label{Ratio_R1}
\end{equation}
is independent of $V$ and takes the universal value
\begin{equation}
  R_1^{(M)}=\frac{e^3}{4h}
  \label{Ratio_R1_univ}
\end{equation}
for bias voltages satisfying Eq. (\ref{Quantum_transport_regime}).
\begin{figure}
\includegraphics[width=8.0 cm]{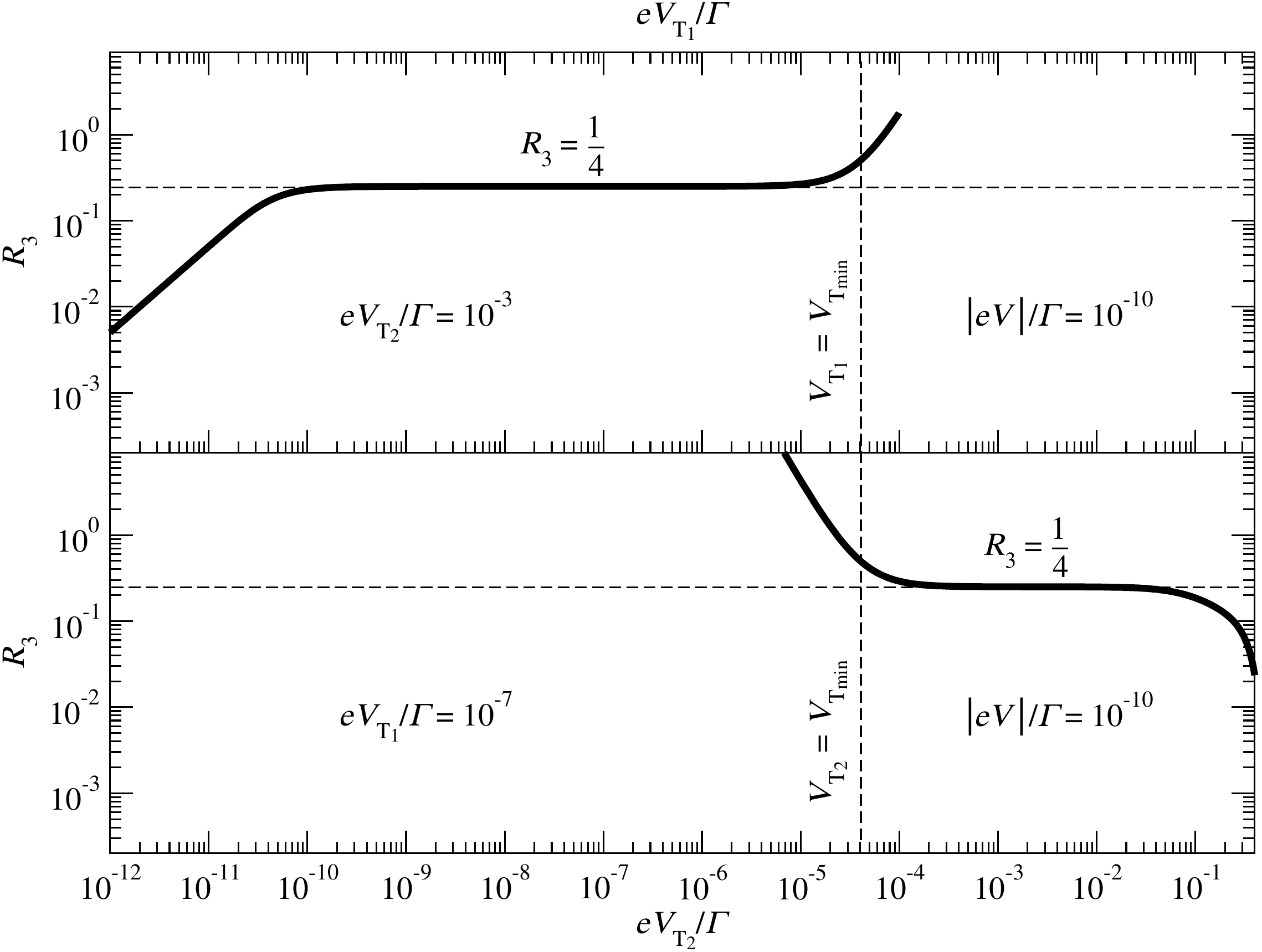}
\caption{\label{figure_3} Dimensionless ratio $R_3$, defined in
  Eq. (\ref{Ratio_R3}), which relates the thermoelectric ($V_{T_1}$ is on the
  left side of the crossover) and pure thermal ($V_{T_2}$ is on the right side
  of the crossover) branches of the differential shot noise
  $\partial S^>/\partial V$ via its value at the crossover
  $V_T=V_{T,\text{min}}$. Upper (Lower) panel: $R_3$ as a function of
  $V_{T_1}$ ($V_{T_2}$) at a fixed value of $V_{T_2}$ ($V_{T_1}$). Here the
  values of the parameters are as follows: $\epsilon_d/\Gamma=10^{-1}$,
  $|eV|/\Gamma=10^{-10}$, $k_BT/\Gamma=10^{-12}$, $|\eta|/\Gamma=1$,
  $\xi/\Gamma=10^{-14}$, and $eV_{T_1}/\Gamma=10^{-7}$ (lower panel),
  $eV_{T_2}/\Gamma=10^{-3}$ (upper panel). For the above values of the
  parameters one gets from Eq. (\ref{V_T_min}) that
  $eV_{T,\text{min}}/\Gamma\approx 4\cdot 10^{-5}$ (see the vertical dashed
  line).}
\end{figure}

For the mean current at low bias voltages in Ref. \cite{Smirnov_2020a} it has
been found that
\begin{equation}
  \frac{\partial I(V,V_T)}{\partial V_T}=\frac{e^2}{h}\frac{\pi^2}{12}\frac{\epsilon_d(eV_T)}{|\eta|^2}.
  \label{Diff_therm_cond}
\end{equation}
Using Eqs. (\ref{V_T_min}), (\ref{Diff_SN_V_T_min}) and
(\ref{Diff_therm_cond}) one finds that the dimensionless ratio
\begin{equation}
  R_2\equiv\frac{\frac{\partial S^>(V,V_T)}{\partial V}\bigg|_{V_T=V_{T,\text{min}}}}{eV_{T,\text{min}}\frac{\partial^2I(V,V_T)}{\partial V_T^2}}
  \label{Ratio_R2}
\end{equation}
becomes universal,
\begin{equation}
  R_2^{(M)}=[1-\ln(2)]\frac{6}{\pi^2}
  \label{Ratio_R2_univ}
\end{equation}
under the conditions in Eq. (\ref{Quantum_transport_regime}).

If one takes two values of the thermal voltage, $V_{T_1}$ and $V_{T_2}$, such
that $V_{T_1}$ belongs to the thermoelectric branch and $V_{T_2}$ belongs to
the pure thermal branch, then from Eqs. (\ref{Thermoelectric_branch}),
(\ref{Pure_thermal_branch}) and (\ref{Diff_SN_V_T_min}) it follows that the
dimensionless ratio
\begin{equation}
  R_3\equiv\frac{V_{T_1}}{V_{T_2}}\frac{\frac{\partial S^>(V,V_T)}{\partial V}\bigg|_{V_T=V_{T_1}}\frac{\partial S^>(V,V_T)}{\partial V}\bigg|_{V_T=V_{T_2}}}
  {\biggl(\frac{\partial S^>(V,V_T)}{\partial V}\bigg|_{V_T=V_{T,\text{min}}}\biggr)^2}
  \label{Ratio_R3}
\end{equation}
is universal,
\begin{equation}
  R_3^{(M)}=\frac{1}{4},
  \label{Ratio_R3_univ}
\end{equation}
that is independent of $V_{T_1}$ ($V_{T_2}$) at fixed $V_{T_2}$
($V_{T_1}$). This is demonstrated in Fig. \ref{figure_3} where the fixed
values of $V_{T_1}$ (lower panel) and $V_{T_2}$ (upper panel) are respectively
chosen such that $|eV|\ll eV_{T_1}\ll eV_{T,\text{min}}$ and
$eV_{T,\text{min}}\ll eV_{T_2}\ll \Gamma$. As one can see in the upper panel
\begin{figure}
\includegraphics[width=8.0 cm]{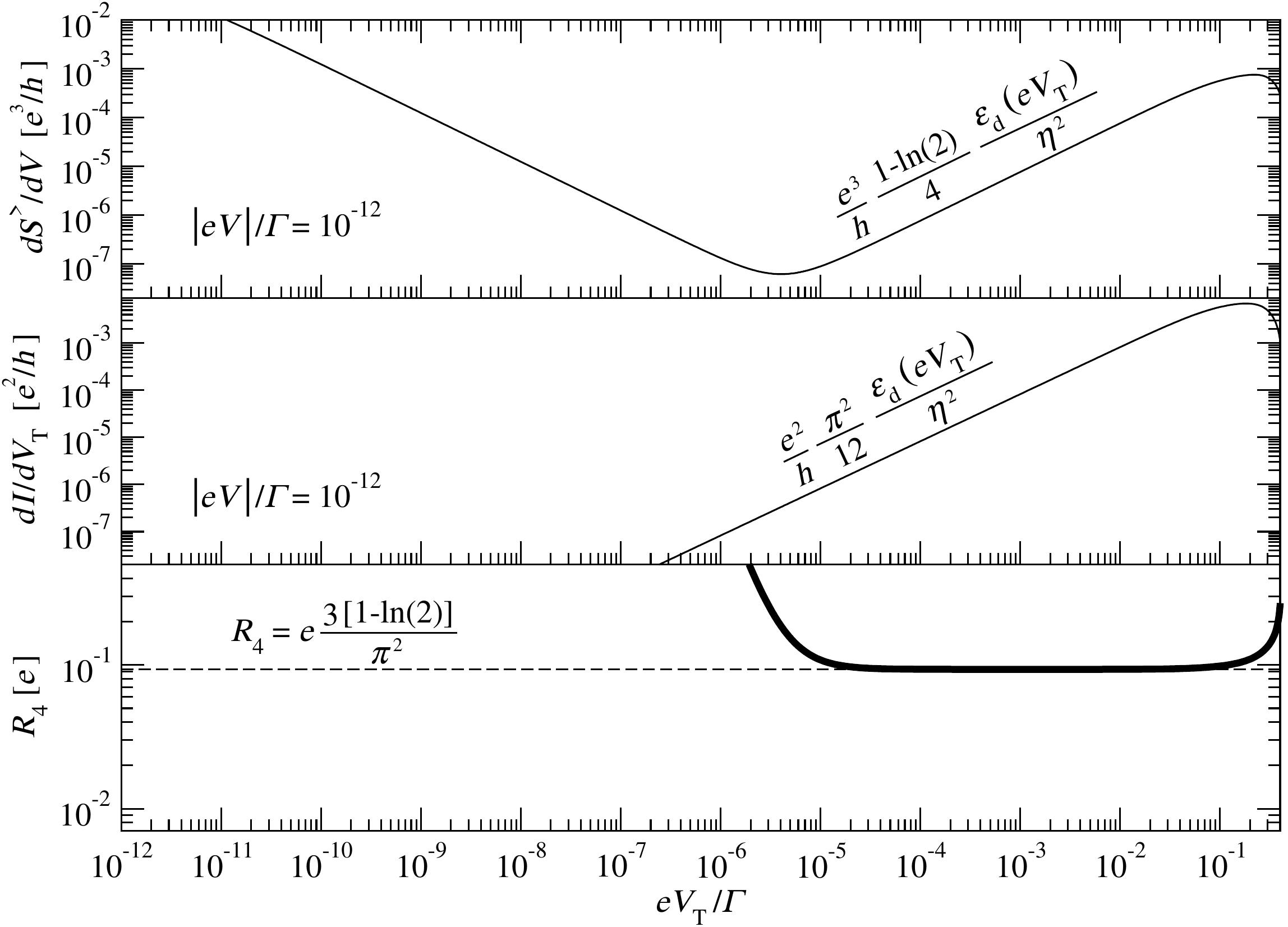}
\caption{\label{figure_4} Differential shot noise $\partial S^>/\partial V$
  (upper panel), differential thermoelectric conductance
  $\partial I/\partial V_T$ (middle panel) and their ratio $R_4$ (lower panel)
  as functions of the thermal voltage $V_T$. Here the results have been
  obtained for the following values of the parameters:
  $\epsilon_d/\Gamma=10^{-1}$, $|eV|/\Gamma=10^{-12}$, $k_BT/\Gamma=10^{-12}$,
  $|\eta|/\Gamma=1$, $\xi/\Gamma=10^{-14}$. According to Eq. (\ref{V_T_min}),
  these values of the parameters give
  $eV_{T,\text{min}}/\Gamma\approx 4\cdot 10^{-6}$.}
\end{figure}
of Fig. \ref{figure_3}, on the left (thermoelectric) side of the crossover the
numerically obtained dimensionless ratio $R_3$ has a perfect plateau with
$R_3=1/4$ (shown by the horizontal dashed line) as expected in the range
$|eV|\ll eV_{T_1}\ll eV_{T,\text{min}}$ for any fixed value of $V_{T_2}$ taken
from the range $eV_{T,\text{min}}\ll eV_{T_2}\ll\Gamma$. As demonstrated in
the lower panel of Fig. \ref{figure_3}, also on the right (pure thermal) side
of the crossover the numerical curve develops a clear plateau on which
$R_3=1/4$ (horizontal dashed line) in the range
$eV_{T,\text{min}}\ll eV_{T_2}\ll \Gamma$ for any fixed value of $V_{T_1}$
taken from the range $|eV|\ll eV_{T_1}\ll eV_{T,\text{min}}$. In contrast,
when $V_{T_1}$ moves away from the thermoelectric branch (left side of the
crossover), that is when $eV_{T_1}\lesssim |eV|$ or
$V_{T_1}\gtrsim V_{T,\text{min}}$, the ratio $R_3$ must deviate from the value
$1/4$. The numerical results (solid curve) in the upper panel show that indeed
deviations from the plateau $R_3=1/4$ occur when $eV_{T_1}\lesssim |eV|$ or
$V_{T_1}\gtrsim V_{T,\text{min}}$. Similarly, when $V_{T_2}$ is not located on
the pure thermal branch (right side of the crossover), that is when
$V_{T_2}\lesssim V_{T,\text{min}}$ or $eV_{T_2}\gtrsim\Gamma$, the ratio $R_3$
must also shift away from the plateau on which it reaches the value $1/4$. As
anticipated, the solid curve resulting from numerical calculations
demonstrates that its plateau-like behavior in the lower panel breaks in the
domains where $V_{T_2}\lesssim V_{T,\text{min}}$ or $eV_{T_2}\gtrsim\Gamma$.

Now let us consider only the pure thermal nonequilibrium branch of the
differential shot noise. From Eqs. (\ref{Pure_thermal_branch}) and
(\ref{Diff_therm_cond}) one finds that the ratio
\begin{equation}
  R_4\equiv\frac{\frac{\partial S^>(V,V_T)}{\partial V}}{\frac{\partial I(V,V_T)}{\partial V_T}}
  \label{Ratio_R4}
\end{equation}
becomes universal,
\begin{equation}
  R_4^{(M)}=e\frac{3[1-\ln(2)]}{\pi^2},
  \label{Ratio_R4_univ}
\end{equation}
in the range of the pure thermal nonequilibrium branch. In Fig. \ref{figure_4}
we present numerical results for the ratio $R_4$.  From the upper and middle
panels one immediately sees the qualitative difference between the behavior of
respectively the differential shot noise $\partial S^>(V,V_T)/\partial V$ and
differential thermoelectric conductance
$\partial I(V,V_T)/\partial V_T$. Indeed, whereas the differential shot noise
$\partial S^>(V,V_T)/\partial V$ exhibits in its minimum a crossover
separating a thermoelectric branch (left side of the crossover) from a pure
thermal branch (right side of the crossover), the differential thermoelectric
conductance $\partial I(V,V_T)/\partial V_T$ does not demonstrate any
crossover and has only one, pure thermal, nonequilibrium branch. As has been
discussed above, in contrast to the universal thermoelectric branch of
$\partial S^>(V,V_T)/\partial V$, its pure thermal branch is not universal
because of its dependence on the gate voltage $\epsilon_d$ and the Majorana
tunneling amplitude $|\eta|$ as can be seen in
Eq. (\ref{Pure_thermal_branch}). Comparing the two non-universal pure thermal
branches of the differential shot noise and differential thermoelectric
conductance, Eqs. (\ref{Pure_thermal_branch}) and (\ref{Diff_therm_cond}),
respectively, one sees that both of them depend linearly on the thermal
voltage $V_T$ and have identical parametric dependence on the gate voltage
$\epsilon_d$ and Majorana tunneling amplitude $|\eta|$. Thus, although the
pure thermal branches of the differential shot noise and differential
thermoelectric conductance are not universal when considered separately from
each other, their ratio $R_4$ in Eq. (\ref{Ratio_R4}), must be universal, that
is it must be independent of the thermal voltage $V_T$, gate voltage
$\epsilon_d$ and Majorana tunneling amplitude $|\eta|$. Moreover, according to
Eq. (\ref{Ratio_R4_univ}), one expects that in the range of $V_T$
corresponding to the pure thermal branch of $\partial S^>(V,V_T)/\partial V$
the ratio $R_4$ must be equal to $3[1-\ln(2)]/\pi^2$ in the universal units of
the elementary charge $e$. The numerical results presented in the lower panel
confirm this expectation: on the right (pure thermal) side of the crossover of
$\partial S^>(V,V_T)/\partial V$ the ratio $R_4$ exhibits a plateau-like
behavior in the range $eV_{T,\text{min}}\ll eV_T\ll\Gamma$ with
$R_4=3e[1-\ln(2)]/\pi^2$ on the plateau. At this point we would also like to
note that, similarly to the differential thermoelectric conductance
$\partial I(V,V_T)/\partial V_T$, the differential conductance
$\partial I(V,V_T)/\partial V$ does not exhibit any crossover. Our numerical
calculations show that it remains almost independent of $V_T$ and retains its
Majorana fractional value $\partial I(V,V_T)/\partial V=e^2/2h$ up to
$eV_T\sim\Gamma$ where it starts to decrease and becomes strongly suppressed,
{\it i.e.} $\partial I(V,V_T)/\partial V\ll e^2/2h$, for
$eV_T\geqslant\Gamma$.
\begin{figure}
\includegraphics[width=8.0 cm]{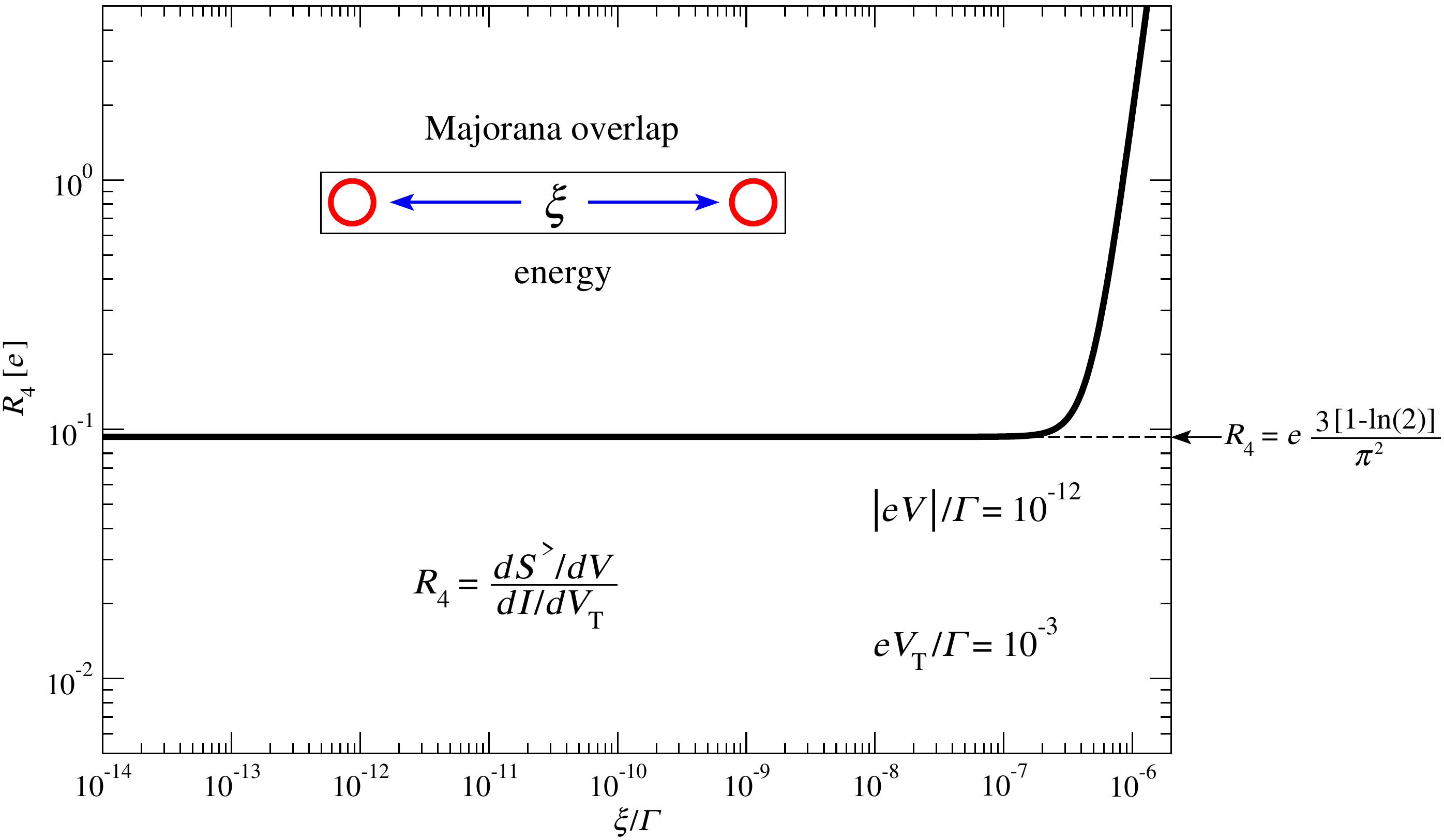}
\caption{\label{figure_5} Ratio $R_4$ defined in Eq. (\ref{Ratio_R4})
  (duplicated also below the solid curve in the figure) as a function of the
  Majorana overlap energy $\xi$. Here we put $\epsilon_d/\Gamma=10^{-1}$,
  $|eV|/\Gamma=10^{-12}$, $eV_T/\Gamma=10^{-3}$, $k_BT/\Gamma=10^{-12}$,
  $|\eta|/\Gamma=1$.}
\end{figure}

To see what happens when the two MBSs are not well separated, we have
performed numerical calculations for larger values of the Majorana overlap
energy $\xi$. Our results show that the above discussed crossover and
universal values of the ratios $R_{1,2,3,4}$ disappear. For example,
Fig. \ref{figure_5} shows the ratio $R_4$ as a function of $\xi$. For well
separated MBSs the values of the parameters are chosen to drive the system
into the regime where it stays  within the plateau shown in the lower panel of
Fig. \ref{figure_4}, that is when the ratio between the pure thermal branch of
the differential shot noise $\partial S^>(V,V_T)/\partial V$ and differential
thermoelectric conductance $\partial I(V,V_T)/\partial V_T$ takes its
universal value, $R_4^{(M)}=3e[1-\ln(2)]/\pi^2$. As Fig. \ref{figure_5}
clearly demonstrates, for small values of the Majorana overlap energy $\xi$
the ratio between the pure thermal nonequilibrium branches of
$\partial S^>(V,V_T)/\partial V$ and $\partial I(V,V_T)/\partial V_T$ is equal
to its universal Majorana value $R_4^{(M)}$. However, when $\xi$ grows, the
two MBSs significantly merge into a single Dirac fermion and cannot be probed
separately anymore. In this situation the universal nonequilibrium Majorana
behavior breaks. As a consequence, the ratio $R_4$ significantly deviates from
its universal Majorana plateau $R_4^{(M)}$. Moreover, for large values of the
Majorana overlap energy $\xi$ both the thermoelectric and pure thermal
nonequilibrium branches of the differential shot noise
$\partial S^>(V,V_T)/\partial V$ are destroyed and the notion of the crossover
discussed above loses its sense as one would expect for a phenomenon having a
Majorana nature.
\begin{figure}
\includegraphics[width=8.0 cm]{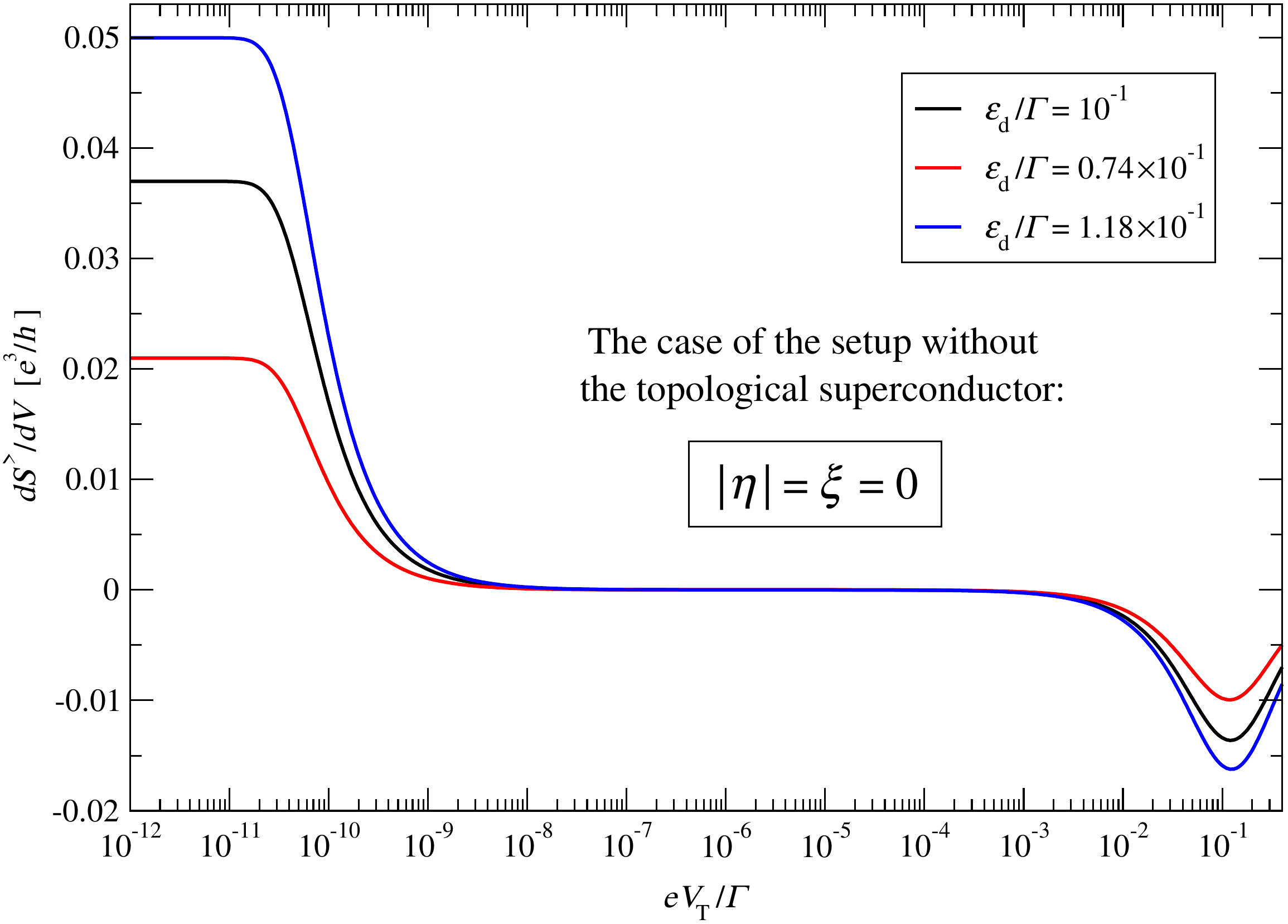}
\caption{\label{figure_6} Differential shot noise $\partial S^>/\partial V$ in
  the absence of MBSs is shown as a function of the thermal voltage $V_T$ for
  three different values of the gate voltage: $\epsilon_d/\Gamma=10^{-1}$
  (black), $\epsilon_d/\Gamma=0.74\times 10^{-1}$ (red),
  $\epsilon_d/\Gamma=1.18\times 10^{-1}$ (blue). The other parameters have the
  following values: $|eV|/\Gamma=10^{-10}$, $k_BT/\Gamma=10^{-12}$,
  $|\eta|=\xi=0$.}
\end{figure}

Finally, to demonstrate that the universal Majorana thermoelectric crossover
in the differential shot noise represents a specific behavior strikingly
distinct from what is observed in conventional systems without coupling to
MBSs, we have computed $\partial S^>(V,V_T)/\partial V$ in the absence of the
topological superconductor. The Hamiltonian of the setup without the
topological superconductor is obtained from our Hamiltonian if one sets
$|\eta|=\xi=0$. The results obtained for  this setup are shown in
Fig. \ref{figure_6}. As can be seen, in the absence of MBSs the dependence of
the differential shot noise on $V_T$ is qualitatively different from the
Majorana induced behavior in two respects.

First, as Fig. \ref{figure_6} shows, the differential shot noise becomes a
monotonic function of $V_T$ for $|eV|\ll eV_T\ll\Gamma$. Its monotonically
decreasing character makes $\partial S^>(V,V_T)/\partial V$ negative at some
point (that is why we avoid using the logarithmic scale for the $y$-axis). In
this sense the differential shot noise is not qualitatively singled out
because the differential electric and thermoelectric conductances are also
monotonically decreasing functions of $V_T$ for $|eV|\ll eV_T\ll\Gamma$ in the
absence of MBSs. Specifically, for $|eV|\ll eV_T\ll\Gamma$ the differential
electric conductance is almost independent of $V_T$ (and is equal to a value
which depends on the gate voltage $\epsilon_d$) up to $eV_T\sim\Gamma$ where
it starts to quickly decrease, whereas the differential thermoelectric
conductance is, unlike the Majorana case, always negative and decreases
linearly with $V_T$ (that is its absolute value grows). In contrast, when MBSs
are present, the differential shot noise is qualitatively singled out by its
nonmonotonic behavior characterized by a minimum, specifying the
thermoelectric crossover, as opposed to the differential electric and
thermoelectric conductances having monotonic behavior exhibiting no minima or
maxima for $|eV|\ll eV_T\ll\Gamma$.

Second, the three curves in Fig. \ref{figure_6}, corresponding to three
different values of the gate voltage $\epsilon_d$, demonstrate that in the
whole range of the thermal voltage $V_T$ the differential shot noise is not
universal, that is in the absence of MBSs $\partial S^>(V,V_T)/\partial V$
depends on $\epsilon_d$ for any value of $V_T$. This nonuniversal behavior of
$\partial S^>(V,V_T)/\partial V$ is qualitatively different from what has been
demonstrated in Fig. \ref{figure_2} where coupling to MBSs makes the
thermoelectric branch (left side of the crossover) of
$\partial S^>(V,V_T)/\partial V$ universal that is independent of
$\epsilon_d$. In contrast, without coupling to MBSs even small variations of
the gate voltage $\epsilon_d$ produce large changes in the differential shot
noise $\partial S^>(V,V_T)/\partial V$ in the whole range of the thermal
voltage $V_T$ as it is clearly seen in Fig. \ref{figure_6}.
\section{Conclusion}\label{concl}
We have explored the differential shot noise $\partial S^>(V,V_T)/\partial V$
in a Majorana entangled QD device driven out of equilibrium by both the bias
voltage $V$ and thermal voltage $V_T$. The numerical analysis of high
precision has been used to reveal the existence of a crossover in the behavior
of $\partial S^>(V,V_T)/\partial V$ as a function of $V_T$ and identify its
analytic form. In particular, it has been shown that this crossover results
from an interplay between the two types of nonequilibrium fluctuations induced
by respectively $V$ and $V_T$ and separates thermoelectric nonequilibrium
behavior of the differential shot noise from its pure thermal nonequilibrium
behavior. The energy scale of the crossover as well as its nonequilibrium
fluctuation nature invisible for mean current probes have been identified and
the crossover dependences on the gate voltage, bias voltage and Majorana
tunneling amplitude have been explicitly shown. Additionally, various
universal Majorana ratios $R_{1,2,3,4}$ involving the energy scale of the
crossover have been provided for a future experimental access to universal
fluctuation behavior of Majorana entangled states within either pure noise
measurements, ratios $R_{1,3}$, or in combination with measurements of mean
currents, ratios $R_{2,4}$. It has been found that the crossover is destroyed
when the two MBSs of the topological superconductor start to overlap and merge
into a single Dirac fermion. This results in a disappearance of the universal
Majorana plateaus in the ratios $R_{1,2,3,4}$ as has been exemplified via
numerical calculations for $R_4$. Finally, we have demonstrated that whereas
for Majorana entangled states the differential shot noise has a nonmonotonic
behavior characterized by a minimum with universal properties, in conventional
systems without coupling to MBSs the differential shot noise is a monotonic
and nonuniversal function in the whole range of $V_T$. Thus, in contrast to
Majorana entangled states, in setups without MBSs the monotonic differential
shot noise is not qualitatively different from the differential electric and
thermoelectric conductances which are also monotonic functions of $V_T$ in the
absence of MBSs.

For an experimental verification of the theoretical results presented in this
work one might consider the devices studied in
Refs. \cite{Deng_2016,Deng_2018}. These devices are based on InAs nanowires
covered by an Al layer grown by molecular beam epitaxy. The Al layer is the
superconductor which is used to induce a topological superconducting state in
the InAs nanowire whose ends are assumed to host MBSs $\gamma_{1,2}$. To
couple $\gamma_1$ to a QD the Al layer is etched on one end of the InAs
nanowire. This bare part of InAs is the place where one forms a QD coupled to
the Majorana state $\gamma_1$ with the coupling strength $|\eta|$. As
explained in Ref. \cite{Deng_2018}, the occupancy (or the energy level
$\epsilon_d$ in our context) of the QD is tuned by proper gate voltages. In
addition to the setup in Refs. \cite{Deng_2016,Deng_2018}, one may also form
two independent normal metallic contacts coupled to the QD with the coupling
strength $\Gamma$. These two independent normal metallic contacts may, in
general, have different chemical potentials $\mu_{L,R}$ and different
temperatures $T_{L,R}$. To measure the differential shot noise one could try
to adapt, for example, the technology from Ref. \cite{Basset_2012} based on
coupling of a setup to a quantum noise detector. Here for the quantum noise
detector one also uses Al as a superconductor and thus it might be compatible
with the above technology \cite{Deng_2016,Deng_2018} for topological
superconductivity. The setup in Ref. \cite{Basset_2012} is a carbon
nanotube. It may be replaced with the InAs nanowire from
Refs. \cite{Deng_2016,Deng_2018}. One possible problem here is that
measurements in Ref. \cite{Basset_2012} assume finite
frequencies. Nevertheless, one may still measure the differential shot noise
if the resonant frequencies in Ref. \cite{Basset_2012} are made smaller than
all the relevant energy scales of our setup. This might be achieved, for
example, by increasing the length of the transmission lines in
Ref. \cite{Basset_2012} or by other relevant techniques.

Among possible outlooks we would like to mention setups with Aharonov-Bohm
fluxes \cite{Zou_2023a} or setups where both MBSs are directly entangled with
a QD whose nonequilibrium states are governed by bias voltages and temperature
differences. Majorana interference effects in such setups will emerge through
the Majorana tunneling phases forming a complex interplay with the two
competing flows induced by respectively $V$ and $V_T$ and the fate of the
Majorana crossover in $\partial S^>(V,V_T)/\partial V$ in this situation is an
interesting and important problem. The results presented in this work have
been obtained assuming that interactions between the Majorana entangled setup
and its environment are sufficiently weak. Under certain circumstances,
however, such interactions may have a significant impact on the shot noise via
corresponding inelastic processes \cite{Krainov_2022} and thus represent a
challenge for future models where MBSs are coupled to an external
environment. Another possibility is to study the differential shot noise in
nonequilibrium setups with poor man's MBSs \cite{Tsintzis_2022,Dvir_2023}
which may arise inside QDs when one fine-tunes parameters of such setups to
locate their states as close as possible to their sweet spots.
\section*{Acknowledgments}
The author thanks Reinhold Egger for valuable comments.

\end{document}